# Strongly correlated excitonic insulator in atomic double layers

Liguo Ma[1], Phuong X. Nguyen[1], Zefang Wang[1], Yongxin Zeng[2], Kenji Watanabe[3], Takashi Taniguchi[3], Allan H. MacDonald[2], Kin Fai Mak[1,4,5]\*, Jie Shan[1,4,5]\*

[1]School of Applied and Engineering Physics, Cornell University, Ithaca, NY, USA
[2]Department of Physics, University of Texas at Austin, Austin, TX, USA
[3]National Institute for Materials Science, Tsukuba, Japan
[4]Laboratory of Atomic and Solid State Physics, Cornell University, Ithaca, NY, USA
[5]Kavli Institute at Cornell for Nanoscale Science, Ithaca, NY, USA

\*Email: jie.shan@cornell.edu; kinfai.mak@cornell.edu
These authors contributed equally: Liguo Ma and Phuong X. Nguyen.

**Excitonic insulators (EI) arise from the formation of bound electron-hole pairs (excitons) [1,2] in semiconductors and provide a solid-state platform for quantum many-boson physics [3-8]. Strong exciton-exciton repulsion is expected to stabilize condensed superfluid and crystalline phases by suppressing both density and phase fluctuations [8-11]. Although spectroscopic signatures of EIs have been reported [6,12-14], conclusive evidence for strongly correlated EI states has remained elusive. Here, we demonstrate a strongly correlated spatially indirect two-dimensional (2D) EI ground state formed in transition metal dichalcogenide (TMD) semiconductor double layers. An equilibrium interlayer exciton fluid is formed when the bias voltage applied between the two electrically isolated TMD layers, is tuned to a range that populates bound electron-hole pairs, but not free electrons or holes [15-17]. Capacitance measurements show that the fluid is exciton-compressible but charge-incompressible – direct thermodynamic evidence of the EI. The fluid is also strongly correlated with a dimensionless exciton coupling constant exceeding 10. We further construct an exciton phase diagram that reveals both the Mott transition and interaction-stabilized quasi-condensation. Our experiment paves the path for realizing the exotic quantum phases of excitons [8], as well as multi-terminal exciton circuitry for applications [18-20].**

In bulk materials EIs can occur in small band gap semiconductors and small band overlap semimetals [21]. In the semiconductor limit, EIs occur when the electron-hole binding energy of an exciton exceeds the charge band gap. The ground state exciton population is then determined by balancing the negative exciton formation energy against the mean exciton-exciton repulsion energy. Although the concept has been understood for sixty years, it has been challenging to establish distinct experimental signatures of its realization. One problem is that exciton coherence in condensed phases inevitably couples to the crystal Hamiltonian, disordered or not, so that condensation does not imply superfluidity [22]. A second problem is that the exciton population of a particular material depends very sensitively on band structure details and cannot be controlled. Here we solve both problems by establishing electrical control of the chemical potentials of excitons that are spatially indirect; the electron and hole wavefunctions do not interfere so that macroscopic phase coherence is spontaneous, allowing exciton superfluidity [15, 16, 23, [24]. We achieve these properties in atomic 2D semiconductor double layers, whose emergence has opened new paths to realize and control the many-exciton states [15-17, 23-32]. Both the formation of dipolar excitons (i.e. excitons with a permanent dipole) and the reduced dielectric screening of the electrostatic interactions favor strong exciton-exciton repulsion. Separate electrical contacts to isolated electron and hole layers provides a reservoir for interlayer excitons with a conveniently tunable chemical potential that allows us to probe thermodynamic properties of the exciton fluid [15, 16, 24].

Consider TMD double layers with a type II band alignment and a spatially indirect band gap $E_G$ (Fig. 1a). The conduction and valence band edges are located in two separate layers; it is therefore possible to make separate contact to each layer. The application of an interlayer bias voltage, $V_b$, splits the electron and hole electrochemical potentials. It reduces the effective band gap (or the charge gap) by $eV_b$ (here $e$ denotes the elementary charge). When the charge gap, $E_G - eV_b$, is below the binding energy, $E_B$, of an interlayer exciton, formation of an interlayer exciton fluid is expected [15, 16, 24]. The exciton chemical potential is defined by $\mu_X = eV_b$ in thermal equilibrium since it is the sum of the electron and hole electrochemical potentials [15, 16, 24]. Because of the large exciton binding energies in atomically thin TMD semiconductors (10's – 100's meV) [23, 24, 32], exciton fluids with a widely tunable chemical potential are achievable. A reservoir for interlayer excitons is established.

TMD double layer devices with separate electrical contacts have been demonstrated recently [27, 28]. However, equilibrium interlayer exciton fluids whose thermodynamic properties could be studied have not been achieved previously. The earlier devices allowed substantial tunneling between the two layers and therefore operated only in the small tunnel junction resistance regime (that is, electrical contact resistances exceeded the tunnel junction resistance); the electron/hole layers could not be maintained at the same electrochemical potential as the contacts [16]. The new results employ a device design that allows us to reach the large tunnel junction resistance regime and demonstrate direct thermodynamic evidence of an EI ground state.

Figure 1b is a schematic cross-section of our devices (see Methods for details on device design and fabrication). We choose monolayer MoSe$_2$ (Mo) as the electron layer and monolayer WSe$_2$ (W) as the hole layer. A hexagonal boron nitride (hBN) tunnel barrier separates the two TMDs, which are angle-misaligned and have negligible moiré effect. The structure is completed with symmetric top and bottom gates that consist of hBN gate dielectrics (~ 10 nm) and graphite gate electrodes. A key part of the strategy used to achieve small contact resistances is to separate the device into a contact region and a region of interest. We electrostatically dope the TMDs heavily in the contact region, but keep the region of interest charge neutral by employing different barrier thicknesses: 1-2 nm in the region of interest and ~ 10 nm in the contact region. When the device is subjected to anti-symmetric gating, $\Delta \equiv (V_{bg} - V_{tg})/2$, where $V_{bg}$ and $V_{tg}$ are the bottom and top gate voltages, respectively, a vertical electric field is generated. The field induces a large potential drop on the double layers in the contact region, which closes the band gap and introduces free carriers; the field reduces the band gap energy (but does not close it) in the region of interest. The reduced gap energy also favors a smaller bias threshold to form excitons. We achieve good electrical contacts down to ~ 10-15 K with $\Delta = 4.6$ V. Unless otherwise specified, all results presented below are obtained at 15 K.

We characterize the thermodynamic properties of the double layer, including both the exciton compressibility and the charge compressibility, using the capacitance technique (Methods). The penetration capacitance per unit area, $C_P = e \frac{dn_{bg}}{dV_{tg}}$, measures the differential charge density on the bottom gate, $dn_{bg}$, induced by a small AC voltage, $dV_{tg}$, on the top gate (Fig. 1c). It characterizes how well the double layer as a whole screens an AC electric field; it approximately measures the charge compressibility of the entire double layer (Methods).

Figure 1d shows $C_P$ as a function of symmetric gating, $V_g \equiv (V_{tg} + V_{bg})/2$. Symmetric gating (referred to as gating below) shifts the chemical potential of the electrons and holes together; it controls the electron-hole density imbalance in the double layer. We first consider the case of zero bias; the electrons and holes have the same chemical potential. When the chemical potential is inside the band gap of the double layer, the double layer is charge incompressible; $C_P$ is given by the gate-to-gate geometrical capacitance, $C_{gg}$. Once the chemical potential touches either the conduction or the valence band edge, the double layer becomes charge-compressible; it screens out the AC electric field, and $C_P$ drops to a small value. (The observed negative values of $C_P$ arise from the electron correlation effects in the double layer [33] and the non-overlapped monolayer regions in the dual-gated device, Extended Data Fig. 1.) The gate voltage between the rising and falling edges of $C_P$ gives an estimate of the band gap energy $E_G \approx 0.68$ eV (see Extended Data Fig. 2 for $E_G$ as a function of $\Delta$). Figure 1d also shows that the charge gap energy decreases with the application of a bias voltage; the gap closes at $eV_b \approx E_G$. More rigorously, the charge gap energy (that is, the charge chemical potential jump across the incompressible regime) can be evaluated by integrating the penetration capacitance with respect to gate voltage [33], $\Delta\mu \approx \int dV_g \, (C_P/C_{gg})$.

Simultaneously we measure the exciton compressibility through the interlayer capacitance. We probe the differential charge density on one layer (W-layer) induced by an AC voltage $dV_b$ that is applied to the other layer (Mo-layer) (Fig. 1e). The differential charge density, arising from transfer of a charge carrier from one layer to the other, is also the differential density of excitons, $dn_X$. The AC voltage $dV_b$ modulates the exciton chemical potential [15, 16]. The interlayer capacitance per unit area, $C_I = e \frac{dn_X}{dV_b}$, thus directly measures the isothermal exciton compressibility $\kappa_X = \left(\frac{\partial n_X}{\partial \mu_X}\right)_T$. Figure 1f shows the bias dependence of $C_I$ at equal electron and hole densities ($V_g \approx 0$ V). The interlayer capacitance rises abruptly from zero when $V_b$ exceeds a threshold ($\approx E_G/e$). It decreases gradually and approaches the geometrical capacitance of the double layer, $C_{2L}$, with further increase of bias voltage. We determine the exciton density by integrating $C_I$ with respect to the bias voltage, $n_X = \int dV_b \frac{C_I}{e}$ (red curve). An exciton density up to a few times of $10^{12}$ cm$^{-2}$ can be achieved with a bias on the order of 1 V.

We analyze the dual-gated double-layer device based on the parallel-plate capacitor model and known device parameters (Methods). The W-layer is grounded, as in the experiment. Figure 2a shows four doping regions for the double layer as a function of bias and gate voltages when electron-hole interactions are neglected. Here, *i*, *p* and *n* denote intrinsic, positively doped, and negatively doped regions for each layer, respectively. Under small bias and gate voltages, both layers are intrinsic (the *ii*-region). The double layer enters the *pn*-region when $V_b$ exceeds a threshold. The threshold is weakly dependent on $V_g$ because with $C_{2L} \gg C_{gg}$, the bias voltage is much more efficient than the gate voltage in tuning the charge chemical potentials.

The electrostatics simulation is in good agreement with the experimental capacitance results: the area with large $C_P$ ($\approx C_{gg}$, shown in red in Fig. 2b) corresponds to the *ii*-region, and the area with large $C_I$ (also shown in red, Fig. 2c) corresponds to the *pn*-region. This agreement shows that the electron-hole fluid is in equilibrium with the electrodes, which provide an electrical reservoir for interlayer excitons. The conclusion is further supported by the long interlayer exciton lifetime (~ 1 ms), determined from an independent tunneling measurement (Extended Data Fig. 3).

In the absence of electron-hole interactions in the double layer, charge gap closing (at the tip of the red triangle in Fig. 2b) and electron-hole pair injection (emergence of $C_I$, the red region in Fig. 2c) are expected to occur at the same bias threshold that is defined by the band gap energy $E_G$. A careful examination reveals, however, that the bias threshold for pair injection (horizontal white line II) is about 25 mV lower than the threshold for band gap closing (horizontal white line I). There exists a finite range of the bias and gate voltages (enclosed by the red and white dashed lines), within which a charge gap and electrons and holes are both present. In this regime the electron-hole fluid is a bosonic fluid of bound electron-hole pairs. The presence of excitons in the charge-incompressible region is a hallmark of the EI [15, 16]. The difference between the two bias thresholds also provides an estimate of the interlayer exciton binding energy ~ $25 \pm 5$ meV.

The charge-gap can be viewed as density-dependent exciton binding energy $E_B$ since it corresponds to the chemical potential jump from an exciton fluid with one extra electron to one with an extra hole [16], and is generally expected to vanish when a critical (Mott) density is reached [23]. Our ability to continuously vary the density of the exciton fluid in the double layer allows us to measure $E_B$ as a function of exciton density. We vary the exciton density by tuning $V_b$ according to Fig. 1f. At each bias, we determine the charge gap energy $\Delta\mu$ from the gate dependence of the penetration capacitance (Extended Data Fig. 4). We use the charge gap at 15 K to approximate $E_B$. The binding energy decreases continuously with increasing $n_X$ (Fig. 3). The extrapolated value at zero density, $17.5 \pm 2.5$ meV, is consistent with our previous $E_B$ estimate and corresponds to an exciton Bohr radius of $a_B \sim 7$ nm. The binding energy drops to zero around $n_X \approx 8 \times 10^{11}$ cm$^{-2}$ (the Mott density), above which the EI turns into an electron-hole plasma [23]. We obtain $n_X a_B^2 \sim 0.4$, which agrees reasonably well with the Mott density estimate $n_X a_B^2 \sim 0.3 - 0.7$ (Ref. [34, 35]). We do not observe any abrupt transition from the EI to a charge conductor around the Mott density down to 10-15 K, indicating the possibility of a continuous Mott transition.

The charge gap at a given exciton density also depends on temperature (Fig. 3). It decreases with increasing temperature and vanishes around $T_S$. Figure 4a shows $T_S$ (open squares) as a function of exciton density. The trend is similar to the density dependence of $E_B$ (15 K in Fig. 3). We find $k_B T_S$ ($k_B$ denoting the Boltzmann constant) is about 40% of the exciton binding energy for the entire density range. The temperature $T_S$ thus represents the ionization temperature (or Saha temperature) of the exciton fluid [23]. Above $T_S$, a sizable portion of the excitons is ionized and the double layer becomes charge-compressible ($C_P \approx 0$).

Finally, we construct the density-temperature phase diagram of the excitons at equal electron and hole densities in Fig. 4a. The color represents the magnitude of the interlayer capacitance $C_I$, or equivalently, the exciton compressibility $\kappa_X$. The phase diagram consists of a dome of large exciton compressibility (shown in red). The exciton ionization temperature $T_S$ defines the right crossover boundary of the dome: below $T_S$ the system is an EI, and above it, a mixture of excitons and electron-hole plasma. Figure 4b shows representative horizontal line cuts of Fig. 4a. At a given temperature, $C_I$ increases linearly with density, reaches a peak, and gradually approaches the interlayer geometrical capacitance $C_{2L}$. With increasing temperature, the peak shifts towards higher density while the peak amplitude decreases. We mark the peak location ($T^{**}$) with a dotted line in Fig. 4a. The scale of $T^{**}$ is 5-6 times larger than the degeneracy temperature estimated for non-interacting bosons, $T^* = 2n_X/D$ (dashed line) [23], where $D = \frac{m}{\pi \hbar^2}$ is the exciton density of states in 2D, defined by the exciton mass $m \approx m_0$ and the reduced Planck constant $\hbar$ ($m_0$ is the free electron mass).

The density dependence of the exciton compressibility in Fig. 4b indicates the importance of exciton-exciton interactions [15, 16, 24, 25, 36]. We consider an interacting Bose gas model with a coupling constant $g$ (Ref. [24]) for the system below $T_S$. The product $gn_X$ describes the interaction energy per exciton. The mean-field isothermal compressibility is given by $\kappa_X = \frac{D}{gD + [\exp(n_X/DkT) - 1]^{-1}}$ (Methods). The result reduces to a linear density dependence, $\kappa_X \approx n_X/(gn_X + k_B T)$, in the high-temperature (non-degenerate) limit, and then to a constant, $\kappa_X \approx g^{-1}$, in the low-temperature (degenerate) limit. The value of the dimensionless coupling constant $gD \approx 11$ obtained from this analysis (Extended Data Fig. 5) is in reasonably good agreement with mean-field calculations for interacting bosons in TMD double layers [15, 16, 24]. The compressibility peak in Fig. 4b is, however, in contradiction to mean-field-theory with a density-independent interaction. This behavior suggests that the coupling constant increases with density as the exciton wavefunctions start to overlap. The compressibility peak occurs when the thermal excitation energy is comparable to the interaction energy, $k_B T^{**} \sim gn_X$. The ratio of $T^{**}$ and $T^*$ therefore corresponds to $\sim gD/2 \sim 5 - 6$, consistent with the independent analysis in Extended Data Fig. 5.

The large dimensionless coupling constant $gD$ shows that the exciton fluid is strongly correlated. The strong correlation is expected to suppress exciton density fluctuations [9, 37] and enhance the effective degeneracy temperature from $T^*$ to $\sim gDT^*/2 \sim T^{**}$ (Ref. [10]).

The region bound by $T^{**}$ and $T_s$ in Fig. 4a therefore represents a degenerate exciton fluid with suppressed density fluctuations (i.e. a quasi-condensate [9, 10, 37]). The strong correlation is also expected to suppress phase fluctuations and enhance the exciton superfluid transition temperature [9, 10, 37]. Our results pave the path for future studies on exciton superfluidity and the BEC-BCS crossover in TMD double layers. (BEC and BCS stand for Bose-Einstein condensation and Bardeen-Cooper-Schrieffer, respectively).

**Methods**
**Device design and fabrication**
A schematic of the dual-gated MoSe$_2$/hBN/WSe$_2$ devices is shown in Fig. 1b. The top and bottom gates consist of hBN gate dielectric of 10 - 20 nm in thickness and few-layer graphite gate electrodes. The two gates are typically symmetric. The hBN spacer between the two TMD monolayers has a different thickness in the region of interest (~ 1.5 - 2 nm) and in the contact region (10 - 20 nm). This is crucial to achieve good metal-TMD contacts as discussed below. The design allows us to achieve equilibrium exciton fluids by biasing the two TMD monolayers at ~ 1 V. Before the charge gap is closed, the tunneling current is negligible (Extended Data Fig. 3) and the exciton density is fully determined by the electrostatics (Fig. 2).

This is in contrast to the previously reported devices [27]. There, excitons can be injected only under high bias (~ 5 - 6 V) because there is a large Schottky barrier at the graphite-TMD contacts and a significant portion of the bias voltage drops at the contact [16]. The electron/hole layer cannot be maintained at the same chemical potential as the contacts. The electrical contact resistance dominates the tunnel junction resistance and the devices are in the small tunnel junction resistance regime. A large tunneling current accompanies the injection of excitons. The exciton density is determined by balancing the pumping and recombination rates of excitons [16]. They are non-equilibrium exciton fluids.

The current design circumvents the electrical contact issue by combining several strategies. We use Pt instead of graphite as contact electrodes; Pt has been reported to form good contacts to both MoSe$_2$ and WSe$_2$ monolayers [38, 39]. We heavily dope the TMDs in the contact region (but not in the region of interest) to form good metal-TMD contacts. This is achieved by applying a large vertical electric field (from anti-symmetric gating). The field creates a vertical voltage drop between the two TMD layers that is linearly proportional to the layer separation. We design a substantially thicker hBN spacer in the contact region than in the region of interest so that the interlayer band gap is closed in the contact region while it is only reduced in the region of interest. In addition, to further suppress electron tunneling between the TMD layers, we increase the thickness of the hBN spacer in the region of interest from 2-3 layers as in the reported devices [27] to 5-6 layers. Our devices are in the large tunnel junction resistance regime. We also find that it is crucial to have the electron and hole contacts close by to inject bound pairs at low temperatures.

We fabricate the devices using the reported dry transfer technique [40]. The constituent layers are exfoliated from bulk TMD crystals (HQ Graphene) and other crystals onto Si substrates covered by a 285 nm-thick SiO2 layer. Their thickness is first identified by optical reflection microscopy and subsequently confirmed by atomic force microscopy (AFM). The layers are picked up from substrates consecutively at ~ 50 ºC using a polymer stamp made of a thin layer of polycarbonate on a polypropylene-carbonate-coated polydimethylsiloxane block. The complete stack is then released onto an amorphous quartz substrate with pre-patterned Pt electrodes at 200 ºC. The residual polycarbonate film is removed in chloroform and isopropanol. Amorphous quartz is chosen as substrates for minimal parasitic capacitance background. The background is also temperature independent, which allows capacitance studies over a broad temperature range. Extended Data Figure 1 shows the optical image of a typical device. We have examined multiple devices in this study. The results are highly reproducible. Main results from an additional device (device 2) are shown in Extended Data Fig. 6.

**Capacitance measurements**
We perform two types of capacitance measurements. The first is the penetration capacitance. We apply a small AC voltage $dV_{tg}$ of root mean square (rms) amplitude 5 mV and frequency 423 Hz to the top gate. It generates an AC electric field that can penetrate the AC-grounded TMD double layer and induce a small charge density $dn_{bg}$ on the bottom gate. The induced charge density is measured using a low-temperature integrated capacitance bridge based on a GaAs high-electron-mobility transistor (HEMT) amplifier [41]. The measurement circuit diagram is shown in Extended Data Fig. 7a. The AC voltage $dV_{tg}$ modulates both the charge carrier density $dn$ and the chemical potential $d\mu$ of the entire double layer. Using an equivalent circuit model for the device (Extended Data Fig. 7b) and ignoring a small contribution from the layer polarizability of the double layer [42], we derive the penetration capacitance, $C_P = e\frac{dn_{bg}}{dV_{tg}} \approx \frac{C_{gg}^2}{C_Q/4 + C_{gg}}$, where $C_Q = e^2 dn/d\mu$ is the junction's quantum capacitance, and $C_{gg}$ is the gate-to-gate geometrical capacitance per unit area. The measured $C_P$ is independent of frequency ranging from 77 Hz to 3.3 kHz.

We also measure the interlayer capacitance (the capacitance of the MoSe2/hBN/WSe2 junctions). The equivalent circuit consists of an interlayer capacitance and a tunneling resistance in parallel. For each DC bias voltage and gate voltage, we apply a small AC excitation voltage $dV_b$ (rms amplitude 5 mV and frequency $f$ = 1187 Hz) to the Mo-layer, and detect the induced interlayer current $dI$ from the W-layer. The interlayer current is complex in general. The in-phase component $dI^{(1)}$ corresponds to the interlayer tunneling current; and the out-of-phase component, the displacement current $dI^{(2)}$, is proportional to the change in exciton density $dn_X$. Both in-phase and out-of-phase components (after a preamplifier) are detected with a lock-in amplifier (SR830). The interlayer capacitance reported in the main text is defined by the displacement current, $C_I = \frac{1}{2\pi f A}\frac{dI^{(2)}}{dV_b}$, where $A \approx 34$ μm² is the area of the double layer region. The measurement results are independent of the excitation amplitude (2 - 15 mV) and frequency (77 Hz to 7.7 kHz).

**Interlayer band gap**

The interlayer band gap $E_G$ of the TMD double layer is controlled by anti-symmetric gating $\Delta \equiv (V_{bg} - V_{tg})/2$. It creates an out-of-plane electric field and a voltage drop of $\approx \frac{t_{2L}}{t_{gg}} 2\Delta$ between the TMD layers from the geometrical considerations. Here $t_{2L}$ and $t_{gg}$ are the TMD interlayer separation and the gate-to-gate separation, respectively. The voltage drop reduces the interlayer band gap linearly, $E_G \approx E_G^0 - \frac{t_{2L}}{t_{gg}} 2e\Delta$, where $E_G^0$ is the intrinsic gap value. As described in the main text, the interlayer band gap can be measured from the penetration capacitance under zero bias; it corresponds to the voltage separation between the capacitance rising and falling edges (Fig. 1d). Extended Data Fig. 2 shows the measured gap value as a function of anti-symmetric gating $\Delta$. It is well described by the above linear dependence with a slope of about $0.21e$. This is fully consistent with the geometry of the device $2\frac{t_{2L}}{t_{gg}} \approx 2\frac{2\,nm}{20\,nm} \approx 0.2$. The extrapolated gap value at zero bias (1.6 eV) is also in good agreement with the reported interlayer band gap (under zero electric field) for MoSe$_2$/WSe$_2$ double layers [43]. All the data presented in the main text were taken at $\Delta = 4.6$ V, which corresponds to $E_G \approx 0.68$ eV.

**Electrostatics simulation**
We can model the electrostatics of the devices approximately using a parallel-plate capacitor model (Extended Data Fig. 8) that neglects the exciton binding energy. We express the electron density ($n_e > 0$) in the Mo-layer and the hole density ($n_h > 0$) in the W-layer for given bias voltage $V_b$ and gate voltages $V_{tg}$ and $V_{bg}$ as

$$n_e e^2 \approx 2C_{gg}(eV_{bg} - \phi_M) + C_{2L}(\phi_W - \phi_M), \qquad (1)$$
$$n_h e^2 \approx 2C_{gg}(\phi_W - eV_{tg}) + C_{2L}(\phi_W - \phi_M). \qquad (2)$$

Here $\phi_M = \mu_e - eV_b$ and $\phi_W = -\mu_h$ are the electrostatic potential of the Mo- and W-layer, respectively; $\mu_e > 0$ and $\mu_h > 0$ are the electron and hole chemical potentials, respectively; $C_{gg}$ and $C_{2L}$ are the gate-to-gate and the TMD double layer geometrical capacitance per unit area, respectively. They are determined by the separation between the capacitor plates and the dielectric constant of the dielectric medium. We can relate the chemical potentials to the carrier densities through the electronic density of states, $\mu_e = \frac{E_G}{2} + \frac{\pi \hbar^2}{m_e} n_e$ and $\mu_h = \frac{E_G}{2} + \frac{\pi \hbar^2}{m_h} n_h$. Here $m_e \approx m_h (\approx 0.5 m_0)$ [44] are the electron and hole band masses ($m_0$ denoting the free electron mass), and $\hbar$ is the reduced Planck constant. We solve equations (1) and (2) for $n_e$ and $n_h$. The result in Fig. 2a corresponds to the reported hBN dielectric constant ($\approx 3.5$) [38] and the measured hBN thickness between the gates ($\approx 20$ nm) and between the TMD layers ($\approx 2$ nm) for device 1.

**Interlayer tunneling current and exciton lifetime**
In addition to the interlayer capacitance, we monitor the DC tunneling current $I$ in the double layer. Extended Data Fig. 3 shows the tunneling current as a function of bias voltage $V_b$ near equal electron and hole densities. The different curves correspond to different anti-symmetric gating $\Delta$, which varies the interlayer band gap energy $E_G$. The current onset is observed approximately when $eV_b \gtrsim E_G$.

The different regimes of tunneling are illustrated in Extended Data Fig. 3b for a fixed anti-symmetric gating $\Delta = 4$ V. The tunneling current remains relatively small over an extended range of bias voltage above the threshold (~ 1 nA or smaller, corresponding to a tunneling current density of no more than 0.03 nA/μm$^2$). The tunneling current density is over four orders of magnitude smaller than that in the devices reported in Ref. [27]. Tunneling in this regime arises from a non-resonant process (left inset). When $V_b$ exceeds ~ 1.7 V (independent of $\Delta$), the tunneling current increases dramatically. The value agrees well with the band gap of monolayer MoSe$_2$ (Ref. [43]). In this regime, tunneling arises from a resonant process (right inset) and a non-equilibrium exciton fluid is formed. The exciton density is determined by balancing the pumping and recombination rates, rather than by electrostatics.

We focus on the equilibrium regime ($V_b < 1.7$ V) in this study. In this regime, the exciton lifetime, $\tau_X$, can be estimated from the tunneling current density and the exciton density: $\tau_X \approx \frac{eAn_X}{I}$. Extended Data Figure 3c shows the bias dependence of $\tau_X$. The exciton lifetime is on the order of millisecond.

**Compressibility of a 2D interacting Bose gas**
The density of a 2D interacting Bose gas is defined by the Bose-Einstein statistics as

$$n_X = \int_0^\infty d\epsilon_X \frac{D}{\exp[(\epsilon_X + gn_X - \mu_X)/k_B T] - 1} = -DT \times \ln[1 - \exp\left(\frac{\mu_X - gn_X}{k_B T}\right)]. \quad (3)$$

Here $D = \frac{m}{\pi \hbar^2}$ is the exciton density of states in 2D with exciton mass $m = m_e + m_h$, $\epsilon_X$ and $\mu_X$ are the exciton kinetic energy and chemical potential, respectively, and $g$ is the exciton-exciton coupling constant. The mean-field isothermal exciton compressibility is evaluated as $\kappa_X = \left(\frac{\partial n_X}{\partial \mu_X}\right)_T$.

In the main text we estimate the exciton-exciton coupling constant $g$ in the low-density limit from the crossover of the compressibility from the high-temperature to the low-temperature behavior. We can also evaluate the coupling constant $g$ by analyzing the compressibility in the high-temperature limit. Extended Data Figure 5 shows the temperature dependence of the inverse interlayer capacitance (symbols) at representative densities in this regime. The data are well described by the mean-field result (solid lines), $C_I^{-1} \propto \kappa_X^{-1} \approx g + \frac{k_B}{n_X} T$. The y-intercept is proportional to $g$, and the x-intercept is given by $-gn_X/k_B$. Extended Data Figure 5b shows the magnitude of the x-intercept as a function of density. We extract $g \approx (2.6 \pm 0.1) \times 10^{-14}$ eVcm$^2$ ($gD \approx 11$) from the slope of the linear fit (solid line). A similar value is also obtained from the y-intercept and the known device area of $A$. The positive sign of $g$ indicates a repulsive interaction. The $g$ value is also in good agreement with a mean-field calculation for interacting bosons in TMD double layers ($g = 3 \times 10^{-14}$ eVcm$^2$) [24].

**Data availability**
The data that support the plots within this paper, and other findings of this study, are available from the corresponding authors upon reasonable request.

**Acknowledgements**
We thank Erich J. Mueller and Cory Dean for fruitful discussions.

**References**
1. Mott, N.F. The transition to the metallic state. *The Philosophical Magazine: A Journal of Theoretical Experimental and Applied Physics* **6**, 287-309 (1961).
2. Jérome, D., Rice, T.M. & Kohn, W. Excitonic Insulator. *Physical Review* **158**, 462-475 (1967).
3. Zhu, X., Littlewood, P.B., Hybertsen, M.S. & Rice, T.M. Exciton Condensate in Semiconductor Quantum Well Structures. *Physical Review Letters* **74**, 1633-1636 (1995).
4. Sun, Z. & Millis, A.J. Topological Charge Pumping in Excitonic Insulators. *Physical Review Letters* **126**, 027601 (2021).
5. Eisenstein, J.P. & MacDonald, A.H. Bose–Einstein condensation of excitons in bilayer electron systems. *Nature* **432**, 691-694 (2004).
6. Kogar, A., Rak, M.S., Vig, S., Husain, A.A., Flicker, F., Joe, Y.I., Venema, L., MacDougall, G.J., Chiang, T.C., Fradkin, E., van Wezel, J. & Abbamonte, P. Signatures of exciton condensation in a transition metal dichalcogenide. *Science* **358**, 1314 (2017).
7. Butov, L.V. Condensation and pattern formation in cold exciton gases in coupled quantum wells. *Journal of Physics: Condensed Matter* **16**, R1577-R1613 (2004).
8. Baranov, M.A., Dalmonte, M., Pupillo, G. & Zoller, P. Condensed Matter Theory of Dipolar Quantum Gases. *Chemical Reviews* **112**, 5012-5061 (2012).
9. Z. Hadzibabic & Dalibard, J. Two-dimensional Bose fluids: An atomic physics perspective. *Rivista del Nuovo Cimento* **34**, 389 (2011).
10. Lozovik, Y.E., Kurbakov, I.L., Astrakharchik, G.E., Boronat, J. & Willander, M. Strong correlation effects in 2D Bose–Einstein condensed dipolar excitons. *Solid State Communications* **144**, 399-404 (2007).
11. Ha, L.-C., Hung, C.-L., Zhang, X., Eismann, U., Tung, S.-K. & Chin, C. Strongly Interacting Two-Dimensional Bose Gases. *Physical Review Letters* **110**, 145302 (2013).
12. Cercellier, H., Monney, C., Clerc, F., Battaglia, C., Despont, L., Garnier, M.G., Beck, H., Aebi, P., Patthey, L., Berger, H. & Forró, L. Evidence for an Excitonic Insulator Phase in 1T-TiSe2. *Physical Review Letters* **99**, 146403 (2007).
13. Seki, K., Wakisaka, Y., Kaneko, T., Toriyama, T., Konishi, T., Sudayama, T., Saini, N.L., Arita, M., Namatame, H., Taniguchi, M., Katayama, N., Nohara, M., Takagi, H., Mizokawa, T. & Ohta, Y. Excitonic Bose-Einstein condensation in Ta2NiSe5. *Physical Review B* **90**, 155116 (2014).

14. Du, L., Li, X., Lou, W., Sullivan, G., Chang, K., Kono, J. & Du, R.-R. Evidence for a topological excitonic insulator in InAs/GaSb bilayers. *Nature Communications* **8**, 1971 (2017).
15. Xie, M. & MacDonald, A.H. Electrical Reservoirs for Bilayer Excitons. *Physical Review Letters* **121**, 067702 (2018).
16. Zeng, Y. & MacDonald, A.H. Electrically controlled two-dimensional electron-hole fluids. *Physical Review B* **102**, 085154 (2020).
17. Burg, G.W., Prasad, N., Kim, K., Taniguchi, T., Watanabe, K., MacDonald, A.H., Register, L.F. & Tutuc, E. Strongly Enhanced Tunneling at Total Charge Neutrality in Double-Bilayer Graphene-WSe2 Heterostructures. *Physical Review Letters* **120**, 177702 (2018).
18. Su, J.-J. & MacDonald, A.H. How to make a bilayer exciton condensate flow. *Nature Physics* **4**, 799-802 (2008).
19. Yu. E. Lozovik & Yudson, V.I. A new mechanism for superconductivity: pairing between spatially separated electrons and holes. *Zh. Eksp. Teor. Fiz.* **71**, 738-753 (1976).
20. Dolcini, F., Rainis, D., Taddei, F., Polini, M., Fazio, R. & MacDonald, A.H. Blockade and Counterflow Supercurrent in Exciton-Condensate Josephson Junctions. *Physical Review Letters* **104**, 027004 (2010).
21. Halperin, B.I. & Rice, T.M. Possible Anomalies at a Semimetal-Semiconductor Transistion. *Reviews of Modern Physics* **40**, 755-766 (1968).
22. Kohn, W. & Sherrington, D. Two Kinds of Bosons and Bose Condensates. *Reviews of Modern Physics* **42**, 1-11 (1970).
23. Fogler, M.M., Butov, L.V. & Novoselov, K.S. High-temperature superfluidity with indirect excitons in van der Waals heterostructures. *Nature Communications* **5**, 4555 (2014).
24. Wu, F.-C., Xue, F. & MacDonald, A.H. Theory of two-dimensional spatially indirect equilibrium exciton condensates. *Physical Review B* **92**, 165121 (2015).
25. Skinner, B., Yu, G.L., Kretinin, A.V., Geim, A.K., Novoselov, K.S. & Shklovskii, B.I. Effect of dielectric response on the quantum capacitance of graphene in a strong magnetic field. *Physical Review B* **88**, 155417 (2013).
26. Zhiyuan Sun, Tatsuya Kaneko, Denis Golež & Millis, A.J. Second order Josephson effect in excitonic insulators. *arXiv:2102.10455* (2021).
27. Wang, Z., Rhodes, D.A., Watanabe, K., Taniguchi, T., Hone, J.C., Shan, J. & Mak, K.F. Evidence of high-temperature exciton condensation in two-dimensional atomic double layers. *Nature* **574**, 76-80 (2019).
28. Jauregui, L.A., Joe, A.Y., Pistunova, K., Wild, D.S., High, A.A., Zhou, Y., Scuri, G., De Greve, K., Sushko, A., Yu, C.-H., Taniguchi, T., Watanabe, K., Needleman, D.J., Lukin, M.D., Park, H. & Kim, P. Electrical control of interlayer exciton dynamics in atomically thin heterostructures. *Science* **366**, 870 (2019).
29. Paik, E.Y., Zhang, L., Burg, G.W., Gogna, R., Tutuc, E. & Deng, H. Interlayer exciton laser of extended spatial coherence in atomically thin heterostructures. *Nature* **576**, 80-84 (2019).
30. Li, J.I.A., Taniguchi, T., Watanabe, K., Hone, J. & Dean, C.R. Excitonic superfluid phase in double bilayer graphene. *Nature Physics* **13**, 751-755 (2017).

31. Liu, X., Watanabe, K., Taniguchi, T., Halperin, B.I. & Kim, P. Quantum Hall drag of exciton condensate in graphene. *Nature Physics* **13**, 746-750 (2017).
32. Rivera, P., Yu, H., Seyler, K.L., Wilson, N.P., Yao, W. & Xu, X. Interlayer valley excitons in heterobilayers of transition metal dichalcogenides. *Nature Nanotechnology* **13**, 1004-1015 (2018).
33. Eisenstein, J.P., Pfeiffer, L.N. & West, K.W. Negative compressibility of interacting two-dimensional electron and quasiparticle gases. *Physical Review Letters* **68**, 674-677 (1992).
34. De Palo, S., Rapisarda, F. & Senatore, G. Excitonic Condensation in a Symmetric Electron-Hole Bilayer. *Physical Review Letters* **88**, 206401 (2002).
35. López Ríos, P., Perali, A., Needs, R.J. & Neilson, D. Evidence from Quantum Monte Carlo Simulations of Large-Gap Superfluidity and BCS-BEC Crossover in Double Electron-Hole Layers. *Physical Review Letters* **120**, 177701 (2018).
36. Skinner, B. & Shklovskii, B.I. Anomalously large capacitance of a plane capacitor with a two-dimensional electron gas. *Physical Review B* **82**, 155111 (2010).
37. Kagan, Y., Kashurnikov, V.A., Krasavin, A.V., Prokof'ev, N.V. & Svistunov, B.V. Quasicondensation in a two-dimensional interacting Bose gas. *Physical Review A* **61**, 043608 (2000).
38. Fallahazad, B., Movva, H.C.P., Kim, K., Larentis, S., Taniguchi, T., Watanabe, K., Banerjee, S.K. & Tutuc, E. Shubnikov--de Haas Oscillations of High-Mobility Holes in Monolayer and Bilayer WSe2: Landau Level Degeneracy, Effective Mass, and Negative Compressibility. *Physical Review Letters* **116**, 086601 (2016).
39. Larentis, S., Movva, H.C.P., Fallahazad, B., Kim, K., Behroozi, A., Taniguchi, T., Watanabe, K., Banerjee, S.K. & Tutuc, E. Large effective mass and interaction-enhanced Zeeman splitting of K-valley electrons in MoSe2. *Physical Review B* **97**, 201407 (2018).
40. Wang, L., Meric, I., Huang, P.Y., Gao, Q., Gao, Y., Tran, H., Taniguchi, T., Watanabe, K., Campos, L.M., Muller, D.A., Guo, J., Kim, P., Hone, J., Shepard, K.L. & Dean, C.R. One-Dimensional Electrical Contact to a Two-Dimensional Material. *Science* **342**, 614 (2013).
41. Ashoori, R.C., Stormer, H.L., Weiner, J.S., Pfeiffer, L.N., Pearton, S.J., Baldwin, K.W. & West, K.W. Single-electron capacitance spectroscopy of discrete quantum levels. *Physical Review Letters* **68**, 3088-3091 (1992).
42. Young, A.F. & Levitov, L.S. Capacitance of graphene bilayer as a probe of layer-specific properties. *Physical Review B* **84**, 085441 (2011).
43. Wilson, N.R., Nguyen, P.V., Seyler, K., Rivera, P., Marsden, A.J., Laker, Z.P.L., Constantinescu, G.C., Kandyba, V., Barinov, A., Hine, N.D.M., Xu, X. & Cobden, D.H. Determination of band offsets, hybridization, and exciton binding in 2D semiconductor heterostructures. *Science Advances* **3**, e1601832 (2017).
44. Mak, K.F. & Shan, J. Photonics and optoelectronics of 2D semiconductor transition metal dichalcogenides. *Nature Photonics* **10**, 216-226 (2016).

# Figures

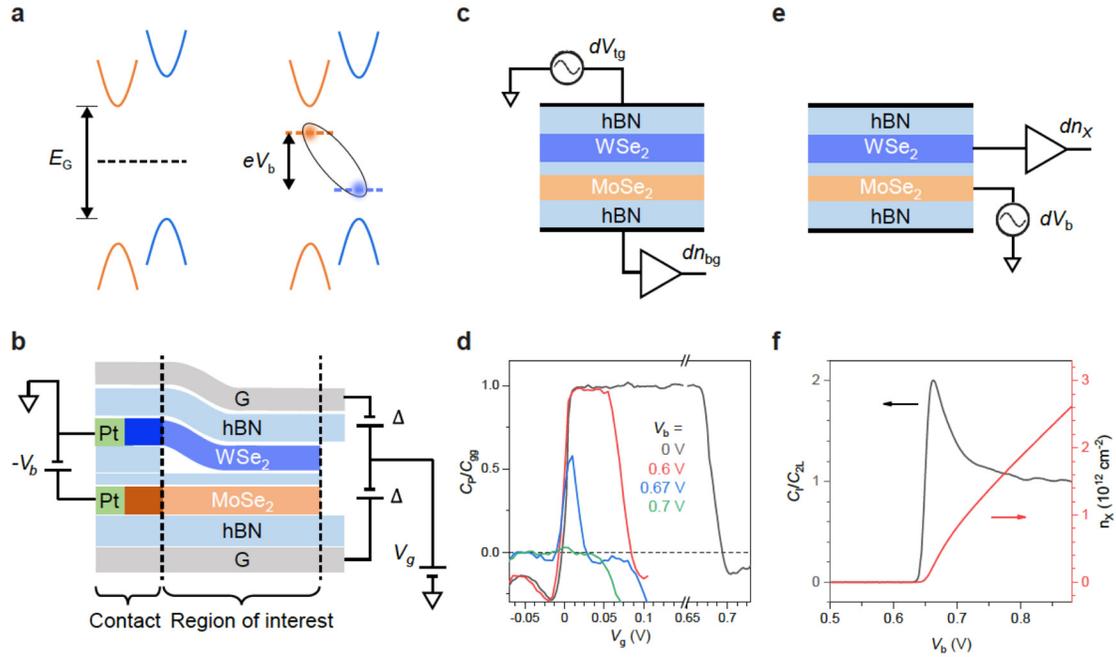

**Figure 1 | Electrical reservoir for interlayer excitons. a,** Type II band alignment of MoSe2/WSe2 double layers with interlayer band gap $E_G$ (left). A bias voltage $V_b$ across the layers separates the electron and hole chemical potentials (dashed lines) and reduces the charge gap (right). Interlayer excitons are formed spontaneously when the bias voltage, which acts as a chemical potential for excitons, exceeds the exciton energy $E_G - E_B$. **b,** Schematic cross-section of dual-gated double layer devices. The anti-symmetric gating $\Delta$ reduces $E_G$ in the region of interest; it closes $E_G$ in the contact region to heavily dope the TMDs. The symmetric gating $V_g$ shifts the electron and hole chemical potentials together and tune their density difference. **c, e,** Schematics of the penetration (**c**) and interlayer capacitance (**e**) measurements. The double layer is AC grounded in **c**. **d,** Gate dependence of the normalized penetration capacitance at representative bias voltages. The sharp capacitance drops signify electron or hole doping into the double layer. The charge gap is closed for $V_b \approx 0.7$ V. **f,** Bias dependence of the normalized interlayer capacitance (black) and the exciton density (red) at $V_g = 0$. **d** and **f** are measured at 15 K and $\Delta = 4.6$ V.

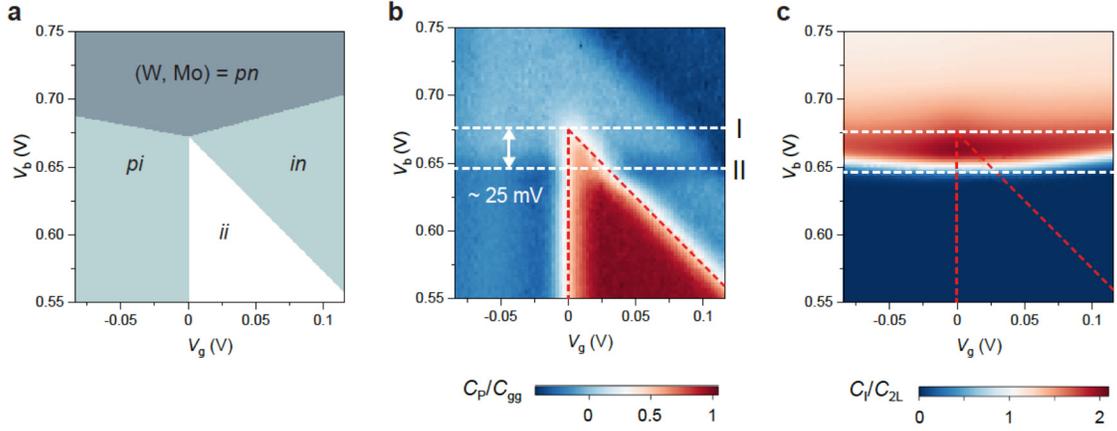

**Figure 2 | Exciton and charge compressibility. a,** Electrostatics simulation of double layer device 1 based on the parallel-plate capacitor model. The electron-hole interaction is ignored. The double layer can be in one of the four doping regions as a function of bias and gate voltages (*i*, *p* and *n* denoting an intrinsic, positively and negatively doped layer). The *ii* region (white) is enclosed by a vertical line (valence band edge of the grounded W-layer) and a line of slope -1 (conduction band edge of the Mo-layer). **b, c,** Normalized penetration capacitance $C_P/C_{gg}$ (**b**) and interlayer capacitance $C_I/C_{2L}$ (**c**) as a function of bias and gate voltages, where $C_{gg}$ and $C_{2L}$ are the gate-to-gate and interlayer geometrical capacitances, respectively. The charge incompressible region in **b** is enclosed by red dashed lines. White dashed line I and II correspond to the bias voltages at which the charge gap closes and the exciton population appears, respectively. Their difference (25 mV) provides an estimate of the exciton binding energy at zero density. The measurements are performed at 15 K and $\Delta$ = 4.6 V.

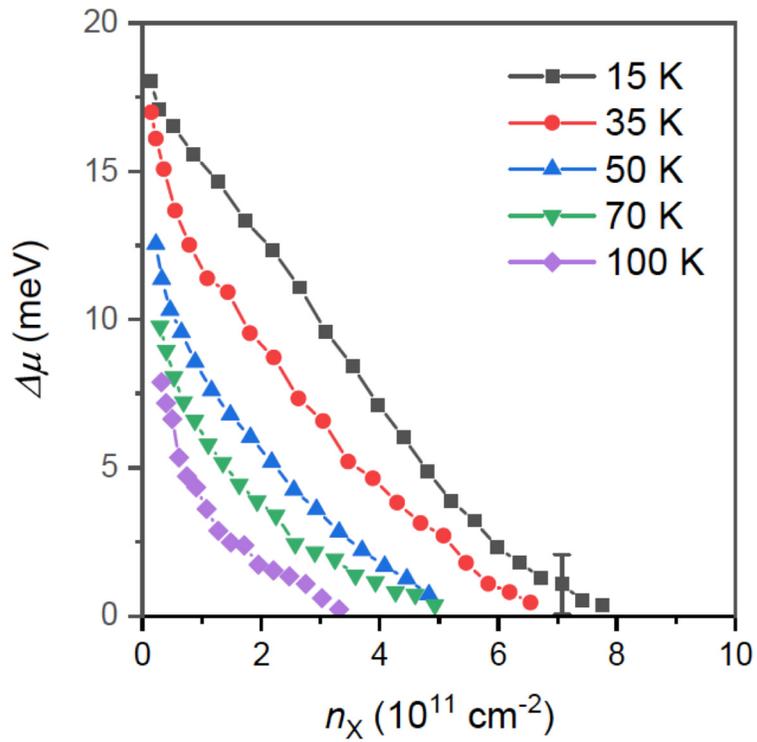

**Figure 3 | Charge gap energy of the double layer.** The charge gap energy as a function of exciton density at different temperatures (symbols). Each data point is extracted from the gate dependence of the penetration capacitance for a fixed bias voltage and temperature. The error bar shows the typical uncertainty of the analysis as described in Extended Data Figure 4. The lines are a guide to the eye.

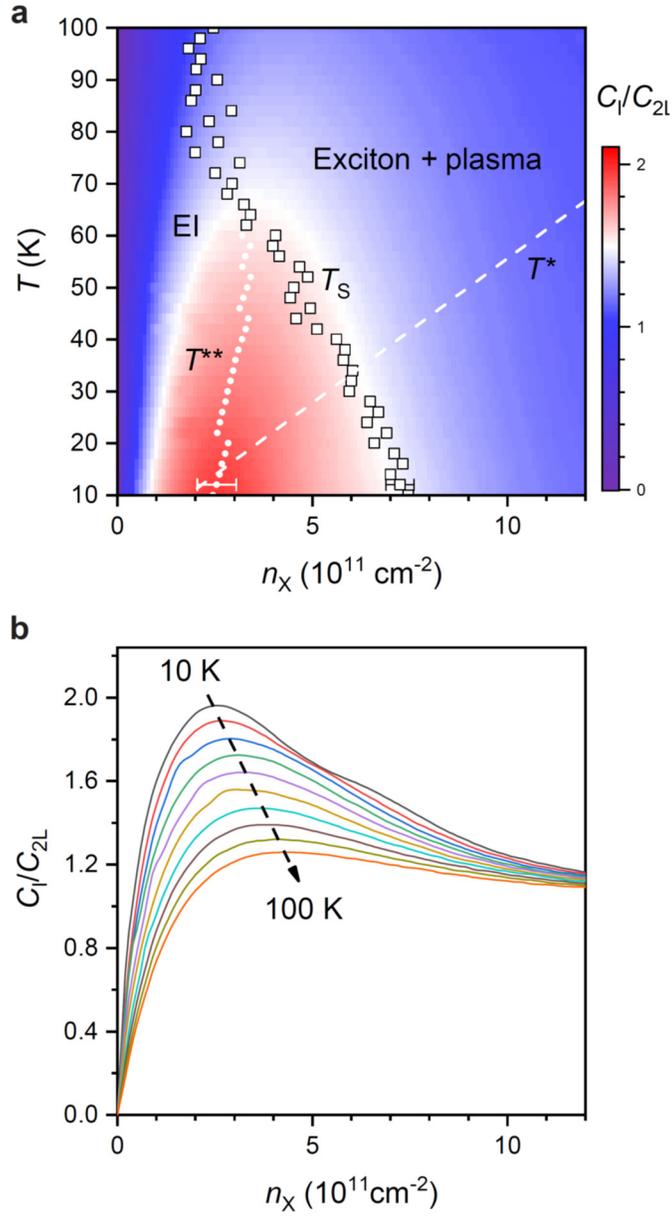

**Figure 4 | Exciton phase diagram. a,** Normalized interlayer capacitance, or equivalently, exciton compressibility as a function of temperature and exciton density. The white squares denote the exciton ionization temperature $T_s$, which separates the crossover from the excitonic insulator (EI) (below) to a mixture of excitons and electron-hole plasma (above). The dashed line $T^*$ and dotted line $T^{**}$ show, respectively, the degeneracy temperature of non-interacting excitons and the temperature for the exciton compressibility peak. The error bar represents the typical uncertainty in density of the compressibility peak. **b,** Exciton compressibility as a function of exciton density at varying temperatures. The compressibility peak divides the high-temperature (left) and low-temperature (right) regime.

**Extended Data Figures.**

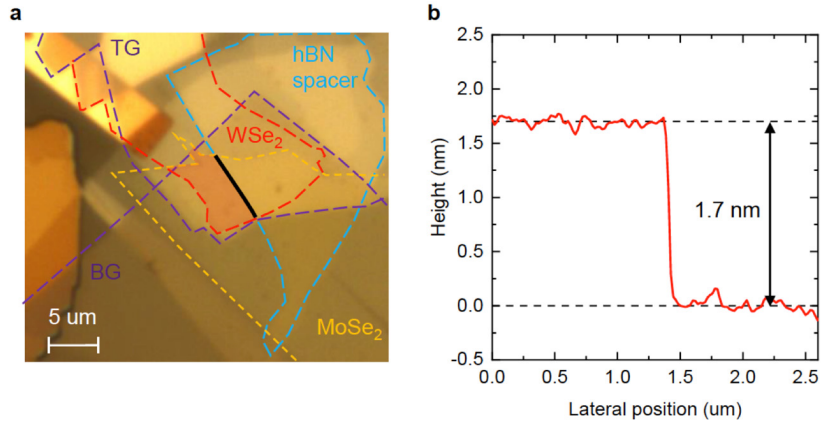

**Extended Data Figure 1 | Device image. a,** Optical image of the MoSe$_2$/WSe$_2$ device presented in the main text (Device 1). The dashed lines mark the boundaries of the top and bottom graphite gates (purple), the MoSe$_2$ monolayer (yellow), the WSe$_2$ monolayer (red), and the thick hBN spacer in the contact region (light-blue). The thin hBN spacer that separates the two TMD layers is not shown. The black solid line marks the boundary of the contact region and the region of interest (shaded). **b,** Atomic force microscopy (AFM) profile of the thin BN spacer, showing a step height of ~1.7 nm (~ 6-layer).

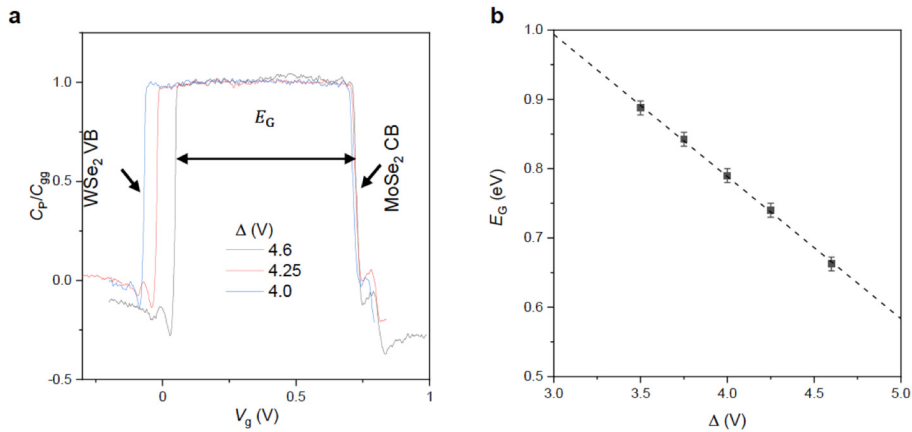

**Extended Data Figure 2 | Dependence of interlayer band gap on anti-symmetric gating. a,** Gate dependence of penetration capacitance at 15 K under varying anti-symmetric gating $\Delta$ and $V_b = 0$ V. The step falls signify electron doping into the MoSe$_2$ conduction band (CB) or hole doping into the WSe$_2$ valence band (VB). The separation between the rising and falling edges determines the band gap $E_G$. An additional step on the electron-doping side arises from the presence of a small isolated MoSe$_2$ monolayer inside the dual-gated device that affects the penetration capacitance. **b,** Interlayer band gap $E_G$ (symbols) extracted from **a** as a function of $\Delta$. The linear fit (dashed line) has a slope of ~ $0.21e$. The gap energy extrapolated for $\Delta = 0$ corresponds to the intrinsic band gap energy $E_G^0 \approx 1.6$ eV.

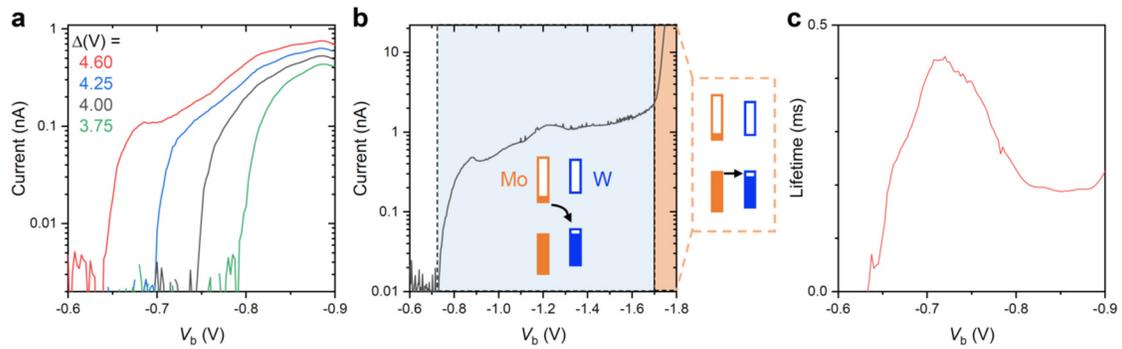

**Extended Data Figure 3 | Tunneling current and exciton lifetime. a,** Bias-dependent interlayer tunneling current of device 1 under $\Delta = 4.6$ V (red), 4.25 V (blue), 4.00 V (black), and 3.75 V (green). The shift of the onset with $\Delta$ is caused by the changing interlayer band gap. **b,** Interlayer tunneling current over a larger bias range at $\Delta = 4$ V. The insets illustrate both the non-resonant (blue-shaded) and resonant (orange-shaded) tunneling regimes. **c,** Estimated exciton lifetime as a function of $V_b$ from the tunneling data at $\Delta = 4.6$ V in **a**.

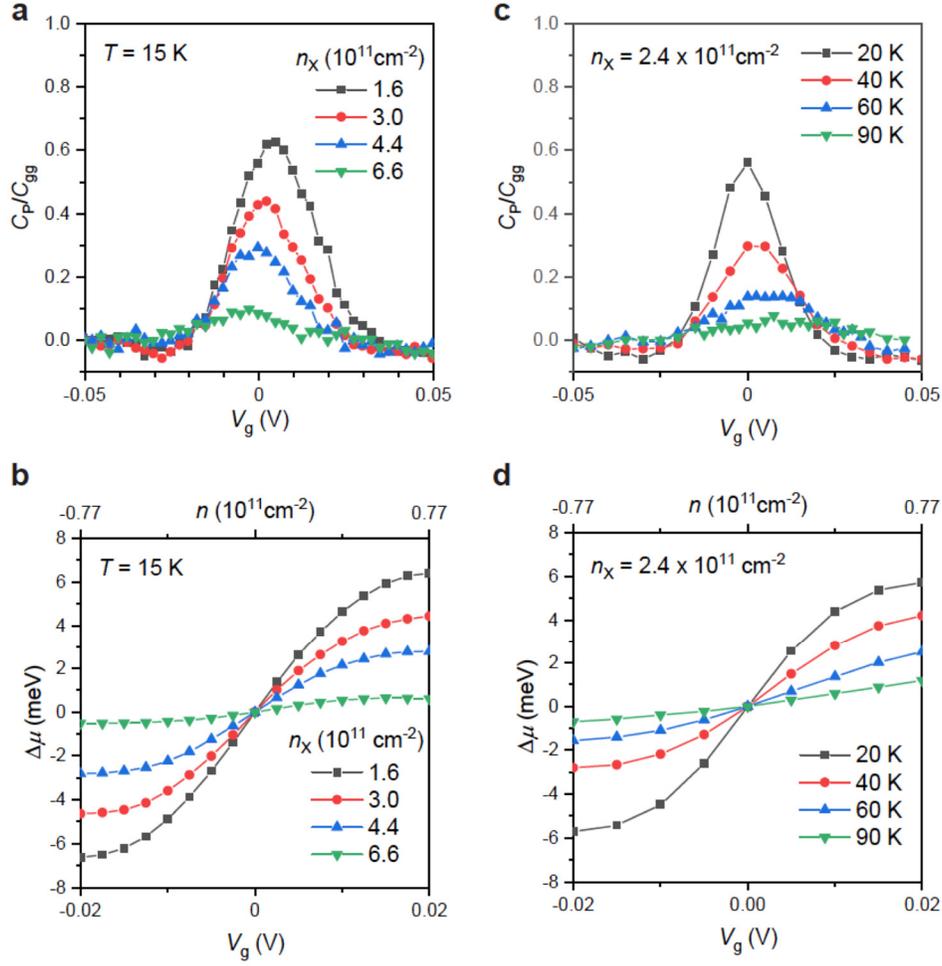

**Extended Data Figure 4 | Determination of the charge gap.** The penetration capacitance (**a**) and the charge chemical potential of the double layer (**b**) at 15 K as a function of $V_g$ at varying exciton densities. The capacitance peak shows the presence of a charge-incompressible state. The integrated area of the peak gives the chemical potential jump (or the charge gap) at equal electron-hole density. The zero point of the chemical potential shift in **b** has been shifted to $V_g = 0$ V for comparison of different exciton densities. The charge gap closes near the Mott density. **c, d,** Similar to **a, b,** at a fixed exciton density $n_X = 2.4 \times 10^{11}$ cm$^{-2}$ for different temperatures. The charge gap closes at the ionization temperature.

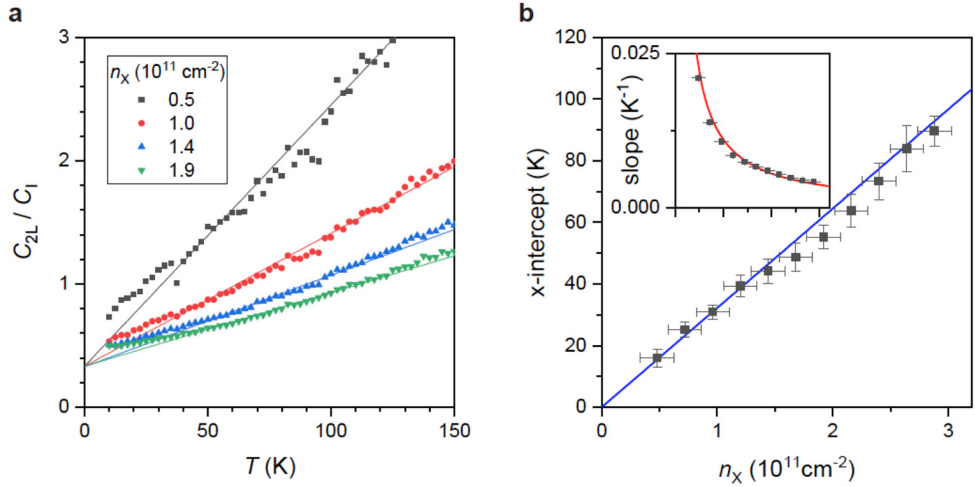

**Extended Data Figure 5 | Exciton compressibility in the high-temperature limit. a,** Inverse interlayer capacitance (or exciton compressibility) as a function of temperature at varying exciton densities. The dashed lines are linear fits in the high-temperature limit, i.e. $gn_X/k_B \lesssim T \lesssim T_s$. **b,** Amplitude of the extracted x-intercept in **a** as a function of exciton density. A linear fit (blue) gives $g = (2.6 \pm 0.1) \times 10^{-14}$ eVcm$^2$. The inset shows the experimental (black) and expected (red) exciton density dependence of the slope of the linear fits in **a**. The density range is the same as the main panel.

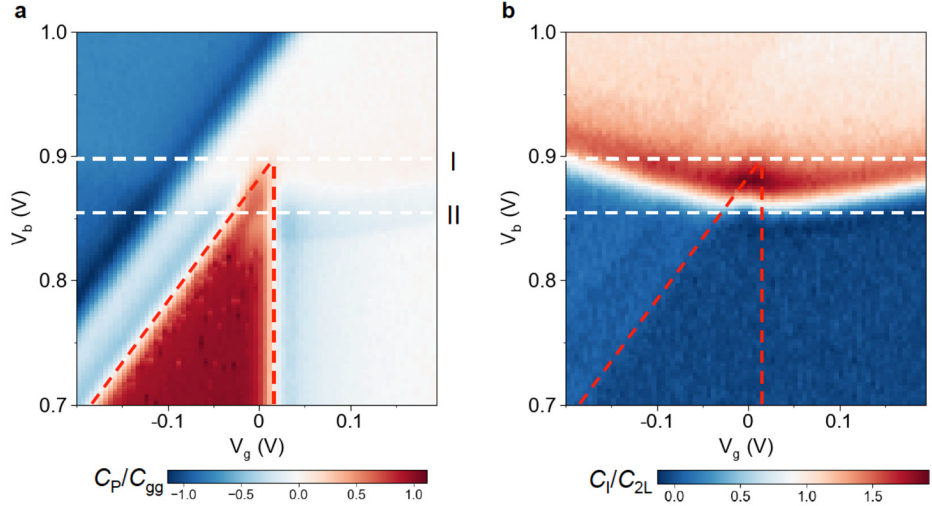

**Extended Data Figure 6 | Main results from device 2.** Penetration capacitance **(a)** and interlayer capacitance **(b)** as a function of bias and gate voltages. The Mo-layer is grounded. The two red dashed lines denote the conduction band edge of MoSe$_2$ (vertical line) and the valance band edge of WSe$_2$ (line with slope +1). The two white dashed lines denote the bias voltage at which the charge gap closes (I) and the exciton fluid becomes compressible (II). The difference between the two values (~ 40 mV) corresponds to the exciton binding energy at $n_X = 0$, which is slightly larger than in Device 1 shown in the main text (~ 25 mV) because the hBN barrier is slightly thinner (~ 5-layer).

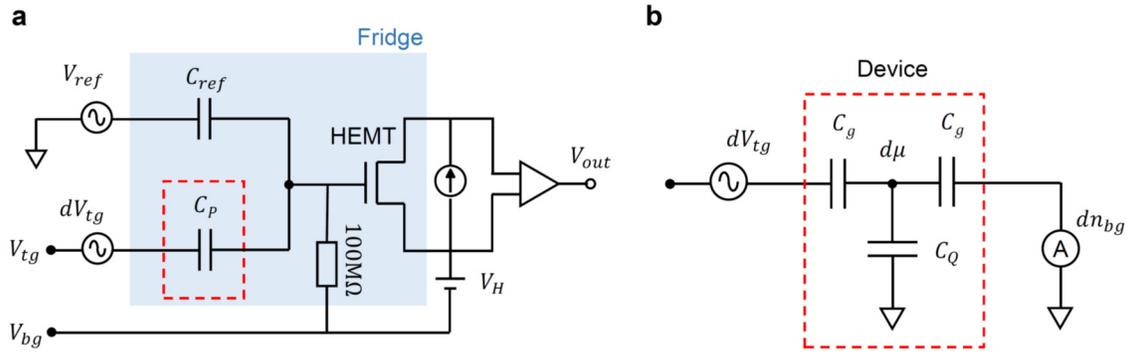

**Extended Data Figure 7 | Circuit diagram. a,** Circuit diagram for the penetration capacitance measurement. The red dashed line encloses the sample area. The reference part ($V_{ref}$ and $C_{ref}$) is used to cancel the parasitic background capacitance. The HEMT is biased at voltage $V_H$. **b,** Equivalent circuit model of $C_P$ in **a**. Here $C_g \approx 2C_{gg}$ is the sample-to-gate geometrical capacitance, which is about twice the gate-to-gate geometrical capacitance $C_{gg}$.

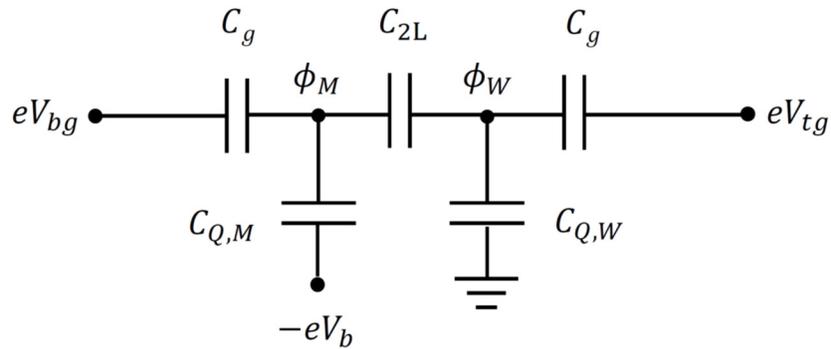

**Extended Data Figure 8 | Equivalent device circuit model for electrostatics simulation.** $C_{Q,M}$ and $C_{Q,W}$ are the quantum capacitances of the MoSe$_2$ and WSe$_2$ monolayers, respectively. Details see Methods.

21